\documentclass{article}
\usepackage{graphicx}
\usepackage{rotating}
\usepackage{amssymb}
\usepackage{mathptmx}
\usepackage[numbers]{natbib}
\usepackage{epstopdf}
\usepackage{authblk}

\makeatletter

\newcommand{\ex}{He$_2^*(a^3\mathrm{\Sigma}^+_u)$}
\renewcommand{\hbar}{\mathchar'26\mkern-9mu h}

\begin{document}

\title{Observation of Crossover from Ballistic to Diffusion Regime for Excimer Molecules in Superfluid $^4$He }

\author[1,2]{D.E.~Zmeev} 
\author[1]{ F.~Papkour}  
\author[1]{P.M.~Walmsley}
\author[1]{A.I.~Golov}
\author[2]{P.V.E.~McClintock}
\author[2]{S.N.~Fisher}
\author[3]{W.~Guo}
\author[3]{D.N.~McKinsey}
\author[4]{G.G.~Ihas}
\author[5]{W.F.~Vinen}

\affil[1]{School of Physics and Astronomy, University of Manchester, Oxford~Road, Manchester M13 9PL, UK}
\affil[2]{Department of Physics, Lancaster University, Lancaster LA1 4YB, UK}
\affil[3]{Department of Physics, Yale University, P.O. Box 208120, New~Haven, CT 06520-8120, USA}
\affil[4]{Department of Physics, University of Florida,\\ P.O. Box 118440, Gainesville, FL 32611-8440, USA}
\affil[5]{School of Physics and Astronomy, University of Birmingham, Birmingham B15 2TT, UK}

\date{07.07.2012}

\maketitle

\begin{abstract} We have measured the temperature dependence of the time of flight of helium excimer molecules \ex~in superfluid $^4$He and find that the molecules behave ballistically
below $\sim$100\,mK and exhibit Brownian motion above $\sim$200\,mK. In the intermediate temperature range the transport cannot be described by either of the models.

\end{abstract}

\section{Introduction}

Einstein's model for Brownian motion suggests that the particle under scrutiny moves ballistically between collisions with the molecules of the surrounding medium. As the medium becomes more rarified,  the free path of the particle increases until the particle becomes ballistic on a macroscopic scale. Superfluid helium at low temperatures behaves as a vacuum that supports a gas of phonons whose density changes with temperature as $T^4$. If one uses a Brownian particle, whose size is much smaller than the wavelength of phonons $2\pi/k$, the particle will experience 
Rayleigh scattering with the cross-section $\sigma \propto \langle k \rangle ^{-4} \propto T^4$. Overall, the free path of such Brownian particle $\ell$ would scale as $T^8$, and its behaviour could be scanned through many orders of magnitude of $\ell$ in a narrow temperature range. This would allow the regime between ballistic motion and diffusion to be studied. 

One candidate for such a Brownian particle is the helium excimer molecule He$_2^*$ in the triplet ($a^3\mathrm{\Sigma}^+_u$) state, which in liquid helium forms a cavity (``bubble'') with a diameter\cite{Eloranta02} of $2a$=1.3\,nm . This is much smaller than the bubbles of 4\,nm in diameter formed around free electrons in helium, and much smaller than the average wavelength of phonons ($2\pi/k\approx$60\,nm at $T=$200\,mK). The molecules form as a result of recombination of positive and negative ions\cite{Tokaryk93}. They were shown to be entrained by the moving normal component in superfluid helium at high temperatures\cite{Mehrotra79}.
Their relatively long half-life\cite{McKinsey99} of  (13$\pm$2)\,s in superfluid is attributed to the forbidden transition to a lower energy state, which requires a spin flip\cite{Shlyapnikov91}.

\section{Experimental details}
\begin{figure}[!h]
\begin{center}
\includegraphics[%
  width=0.8\linewidth,
  keepaspectratio]{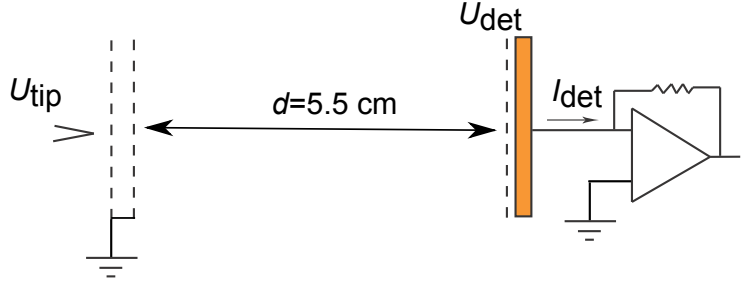}
\end{center}
\caption{Sketch of the experimental setup.}
\label{cell}
\end{figure}

The experimental cell of 0.6\,litres in volume was filled with ultrapure $^4$He (concentration of $^3$He was\cite{Hendry87}  $x_3<5 \times 10^{-13}$) and cooled down with a dilution refrigerator. The sketch of the experimental setup is presented in Fig.~\ref{cell}. The molecules were generated using field emission from a tungsten tip\cite{Ihas71, Golov98} into superfluid helium. Only moderate electric currents of up to $\sim$2nA and short pulses of 50\,ms were used to suppress the formation of turbulent tangles in the superfluid. The corresponding negative tip voltages were up to $U_\mathrm{tip}=700$\,V. Two grounded grids in front of the tip were used to collect electrons. The molecules travelled a distance of $d=5.5$\,cm through the superfluid helium, and their arrival was registered at a detector comprising a copper plate and a stainless steel grid separated by thin (50\,$\mu$m) kapton strips. A voltage of $U_\mathrm{det}$=-1.2\,kV was applied to the grid, so that  the detector electric field was  $\sim2\times 10^5$\,V/cm.  When the molecules collide with a metal surface, they are ionized or knock an electron out of the metal and the electric field separates the opposite charges or the electron from its image. The collected electron current was amplified at room temperature using a SR570 current preamplifier.
The measured time constant of the preamplifier was 5.5\,ms, so the setup was capable of resolving average velocities of up to 10\,m/s.

 The experiments were carried out at two pressures in the cell: $P=$0.1\,bar and $P=$5.0\,bar.

\section{Results}

\begin{figure}[!h]
\begin{center}
\includegraphics[%
  width=0.9\linewidth,
  keepaspectratio]{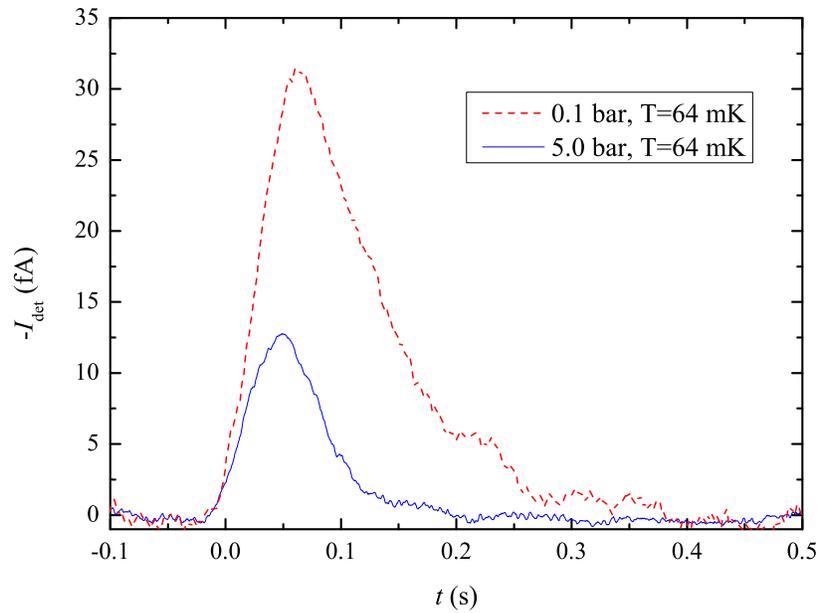}
\end{center}
\caption{Signals obtained at $T=64$\,mK and at two different pressures: 0.1\,bar (dashed red  line) and 5.0\,bar (solid blue line). The tip voltage was pulsed from $t=0$ to 0.05\,s.}
\label{fast}
\end{figure}

Examples of the obtained signals at low temperatures are presented in Fig.~\ref{fast}. For unknown reasons the amplitude of the signals at 5.0\,bar was significantly smaller for a similar tip current.

\begin{figure}[!h]
\begin{center}
\includegraphics[%
  width=0.9\linewidth,
  keepaspectratio]{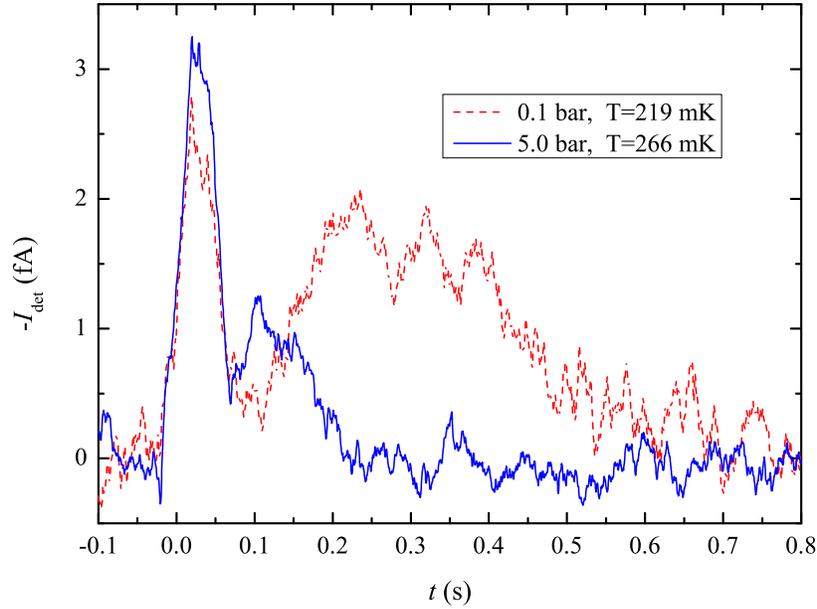}
\end{center}
\caption{(Examples of signals obtained at high temperatures and at two different pressures: 0.1\,bar (dashed red line) and 5.0\,bar (solid blue line). The tip voltage was pulsed from $t=0$ to 0.05\,s. The measured electric tip current was the same as for signals in Fig.~\ref{fast}. The fast peaks probably correspond to photons emitted by short-lived excited helium states. Note  that the vertical scale here is 10 times smaller than in Fig.~\ref{fast}.}
\label{slow}
\end{figure}
\newpage
The time of flight was found to be independent of the amplitude of the tip current, and translated into average velocities of 1.7\,m/s and 2.4\,m/s for the pressures of 0.1 and 5.0\,bar correspondingly. The time of flight here is defined as the time at the maximum of the peak minus the halfwidth of the injection pulse (25\,ms) and minus the time constant of the preamplifier (5.5\,ms). Signals at higher temperatures are shown in Fig.~\ref{slow}. The fast peak on both signals is probably due to photons emitted by atoms in  short-lived excited states, which could lead to  photocurrent. 

\begin{figure}[!h]
\begin{center}
\includegraphics[%
  width=0.78\linewidth,
  keepaspectratio]{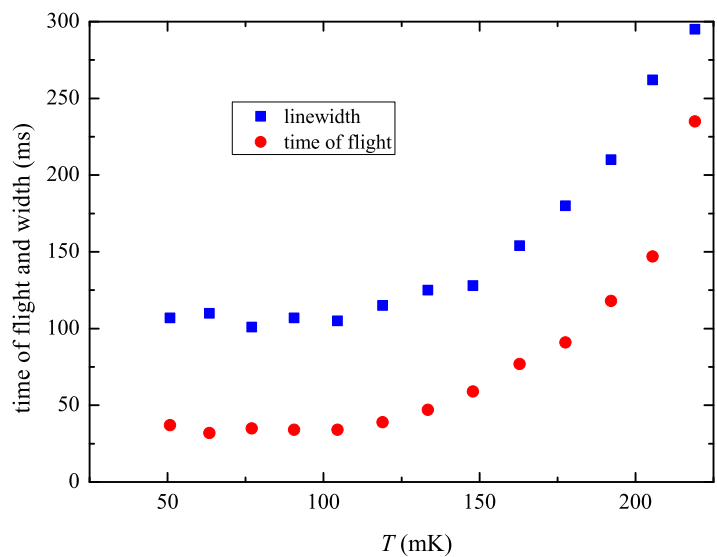}
\end{center}
\caption{Temperature dependence of the time of flight (red circles) and linewidth  of the signals (blue squares). $P=0.1$\,bar. }
\label{tof}
\end{figure}

\begin{figure}[!h]
\begin{center}
\includegraphics[%
  width=0.8\linewidth,
  keepaspectratio]{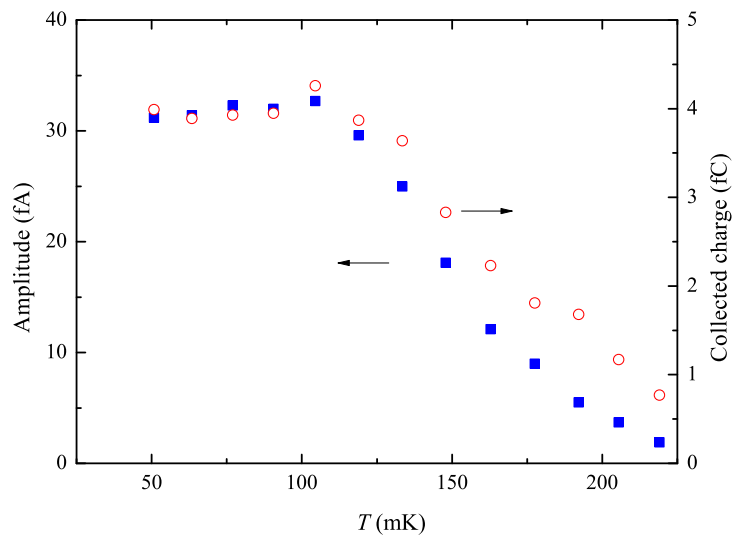}
\end{center}
\caption{Temperature dependence of the amplitude of the signal (blue squares, left-hand axis) and the integral of the signal (collected charge, red circles, right-hand axis). $P=0.1$\,bar.}
\label{ampl}
\end{figure}

The amplitude of the signals and the time of flight did not change up to temperatures of 110 and 150\,mK for signals at 0.1\,bar and 5.0\,bar respectively. Above those temperatures the amplitude started to decrease and the time of flight and the linewidth, defined as the signal width at half the maximum, increased. This is illustrated in more detail in Figs.~\ref{tof}~and~\ref{ampl} for the data obtained at $P=0.1$\,bar.  Similar dependences were observed at $P=$5.0\,bar. 

\newpage

\section{Discussion}

In order to obtain the diffusion coefficient, we recourse to the calculation of mobility of an electron bubble in superfluid $^4$He performed by Baym et al.\cite{Baym69}.
As the radius of the excimer bubble is much smaller than that of the electron bubble, we can neglect the resonant scattering, which is important for phonons of wavelength
comparable to the size of the bubble, and also neglect the compressibility of the bubble. The analogue of mobility for neutral particles is the factor of proportionality between the drift velocity and the force exerted on the excimer:

\begin{equation}
b = \frac{256\pi^7}{405}\hbar a^6\left(\frac{k_\mathrm{B} T}{\hbar c}\right)^8,
\label{bb}
\end{equation}
where $c$ is the speed of sound in helium and $k_\mathrm{B}$ is the Boltzmann constant.
Formula~(\ref{bb}) coincides with the result of Baym et al.\cite{Baym69}  in the limit $T\to 0$. The momentum relaxation time $\tau$ is given by $\tau=m/b$, where $m$ is the effective mass
of an excimer. It equals the sum of the hydrodynamic mass of the bubble and the mass of two $^4$He atoms:   $m = 2\pi/3\rho a^3 + 2 m_4 \approx 13 m_4$,
where $\rho$ is the liquid density. The momentum relaxation distance is of order $\ell=v_0 \tau$, where $v_0$ is the characteristic velocity of an excimer.
The excimers cannot be treated as ballistic particles when $\ell$ becomes comparable to the size of the cell $d=$5.5\,cm. This gives the characteristic temperatures for the onset of the crossover regime as $T=110$\,mK and $140$\,mK for $P=0.1$\,bar and 5.0\,bar correspondingly. Similar temperatures were observed in the experiment.

\begin{figure}
\begin{center}
\includegraphics[%
  width=0.75\linewidth,
  keepaspectratio]{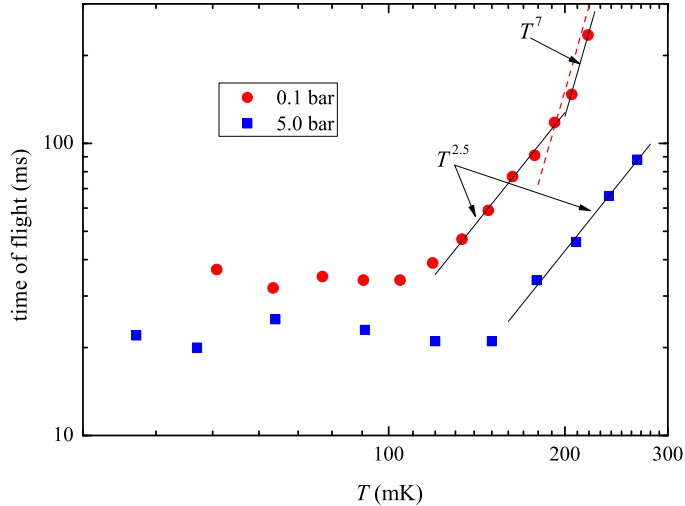}
\end{center}
\caption{ The measured times of flight as a function of temperature at two different pressures $P=0.1$\,bar (red circles) and  $P=5.0$\,bar (blue squares). The  diffusion time (\ref{tdiff}) calculated at $P=$0.1\,bar is shown by the dashed line (no fitting parameters were used). Solid lines show slopes corresponding to the indicated powers of temperature.}
\label{diff}
\end{figure}

The diffusion coefficient is given by $D=k_\mathrm{B} T/b$, so that the mean diffusion time over the distance $d$ is equal to

\begin{equation}
t_\mathrm{diff} =\frac{d^2}{2D} \propto T^7~~.
\label{tdiff}
\end{equation}
This time is calculated without any fitting parameters and compared to the measured time of flight in Fig.~\ref{diff} (dashed line). The diffusion regime at $P=$5.0\,bar failed to  develop in full before the signal disappeared, but the temperature dependences for both pressures in the intermediate regime are consistent with the $T^{2.5}$ power law. We do not have a suitable theory on hand and this fact is yet to be explained.

An interesting fact is that the measured average velocities of the molecules in the ballistic regime were smaller than their calculated thermal velocities $\overline{v_\mathrm{th}}=\sqrt{8k_\mathrm{B} T/ \pi m}$. At $T=50$\,mK the mean thermal velocities of excimers are 4.4\,m/s (4.3\,m/s) for $P$=0.1\,bar (5.0\,bar), while in the experiment mean velocities of the excimers  at low temperatures were 1.7\,m/s (2.4\,m/s) for  $P$=0.1\,bar (5.0\,bar).
Perhaps the excimers do not have time to thermalise after they have been formed (at low temperatures $\ell \gg d$): The most probable mechanism for the creation of molecules is recombination of positive ions and electrons, so that the resulting momentum is the sum of the momenta of the components (which, in turn, were not necessarily thermalised).

In our treatment we did not take into account scattering from $^3$He atoms. In the ballistic regime, the collisions become important when $x_3$ is on the order of 1 atom per volume covered by an excimer trajectory $\pi a^2 d $, or $x_3 \approx 10^{-9}$. As phonon scattering intensifies, the trajectories of excimers become more complicated and the probability of collisions increases. It seems plausible that the excimer molecules cease to exist upon meeting a $^3$He atom. Indeed, the $^3$He atom has a nuclear spin, which provides an interaction that might facilitate the electron spin flip required for the excimer to dissociate. Also,  $^3$He atoms tend to condense on the surfaces of the bubbles, as they decrease the surface tension\cite{Dahm69}. In the case of the electron bubble, the condensation energy\cite{Dahm69} is 0.25\,K per atom and  is probably not very different for the excimer bubble.  In our situation this means that the excimers, which had interacted with $^3$He atoms, did not contribute to the observed signal and interaction with $^3$He impurities can be safely ignored as far as the time of flight is concerned. The same argument applies to excimer-excimer collisions. The annihilation of excimers by $^3$He impurities is consistent with the observed significant decrease in the excimer signal upon introducing $x_3=3\times 10^{-10}$ of $^3$He impurities. On the other hand, excimers were observed in $^4$He with the concentration of $^3$He on the order of $x_3=10^{-7}$ at temperatures above 1\,K\cite{Mehrotra79, McKinsey99}. This could be due to $^3$He impurities not condensing on the bubble surfaces at high temperatures: condensed $^3$He atoms spend longer times and are closer to the excited electron, which makes the spin-spin interaction more effective, and excimers more sensitive to $^3$He impurities at low temperatures.

\section{Conclusions}

We have observed three different kinds of transport of excimer molecules in superfluid $^4$He: ballistic  at low temperatures, diffusion at high temperatures (where the time of flight is proportional to $T^7$), and crossover between the two regimes at intermediate temperatures. We used a primitive analysis to calculate the diffusion coefficient and see a satisfactory agreement with the experiment.  The calculated values for the temperature of crossover onset are consistent with measurements at both experimental pressures $P=$0.1\,bar and $P=$5.0\,bar. The time of flight of excimers in the crossover region was found to be proportional to $T^{2.5}$, for which we do not yet have an explanation.

\section{Acknowledgements}
This work was supported through the Materials World Network program by the Engineering and Physical Sciences Research Council [Grant No. EP/H04762X/1] and the National Science Foundation [Grants DMR-1007937 and  DMR-1007974]  . PMW is indebted to EPSRC for the Career Acceleration Fellowship [Grant No. EP/I003738/1].

\end{document}